\documentclass[aps,prl,reprint]{revtex4-1}
 \usepackage{graphicx}

\begin{document}

\author{Xiaohua Zhang}
\author{Qingwen Li}
\email{liqw05@gmail.com}
\affiliation{Suzhou Institute of Nano-Tech and Nano-Bionics, Ruoshui Road 398, Suzhou 215125, China}

\title{Enhancement of Friction between Carbon Nanotubes: An Efficient Strategy to Strengthen Fibers}

\date{\today}

\begin{abstract}
Interfacial friction plays a crucial role in the mechanical properties of
carbon nanotube based fibers, composites, and devices.  Here we use molecular
dynamics simulation to investigate the pressure effect on the
friction within carbon nanotube bundles.  It reveals that the intertube
frictional force can be increased by a factor of 1.5$\sim$4, depending on tube
chirality and radius, when all tubes collapse above a critical pressure and
when the bundle remains collapsed with unloading down to atmospheric pressure.
Furthermore, the overall cross-sectional area also decreases significantly for
the collapsed structure, making the bundle stronger.  Our study suggests a new
and efficient way to
reinforce nanotube fibers, possibly stronger than carbon fibers, for usage at
ambient conditions.

\end{abstract}

\maketitle

Carbon nanotubes (CNTs) are considered the strongest and ideal reinforcing
fibers due to their exceptional mechanical properties, low density, and high
aspect ratio.  However, although the axial strength and stiffness of individual
CNTs are of the order of 50$\sim$100 GPa and 1 TPa, respectively
\cite{treacy.mmj:1996, lu.jp:1997, yu.mf:20001, ruoff.rs:2003,
dumitrica.t:2006}, the highest
strength of CNT and CNT-reinforced fibers, ranging from 0.85 to 3.3 GPa
\cite{zhang.m:2004, li.qw:2006, zhang.xf:20071, zhang.xf:2007, motta.m:2005,
motta.m:2007, tran.cd:2009}, is nearly 2$\sim$3 orders of magnitude lower than
individual tubes and about 1/3$\sim$1/2 of the strongest Toray carbon fibers \cite{toraynote}.
As good alignment improves the translation of axial properties of individual
tubes to those of the fiber, efforts were reported to grow ultralong and
well-aligned CNT arrays (forests) \cite{li.qw:2006}, and to improve the direct
spinning method \cite{motta.m:2005}.  Post-spin treatments, {\it e.g.},
infiltration, twisting, heating, and stretching, have been reported to improve
the load transfer between CNT bundles, by making better oriented network
\cite{motta.m:2007} and closed packing \cite{tran.cd:2009} of bundles.
However, CNTs often do not exist as individual tubes but group into bundles,
the basic component of the spun fibers.
Therefore it can be of great importance to improve the strength of bundles.  So
far, the bundle strength is reported to be about 10 GPa for the length of
several micrometers \cite{yu.mf:20001}.  Considering the strength loss from
component filaments to traditional fibers \cite{hearle.jws:1967},
it is hard to achieve the same strength as carbon fibers by grouping CNT
bundles through various spinning treatments.  One problem strongly related to
the bundle strength is that individual tubes in bundle tend to slide easily
against each other \cite{salvetat.jp:1999}.  Recently, translational static and
sliding frictions
in a multi-wall CNT were measured to be
0.014 and 0.009 meV/{\AA} per atom, respectively \cite{cumings.j:2000}, and the
frictional force
in a bundle was reported, surprisingly, of several orders of magnitude greater
\cite{syue.sh:2006, yang.ty:2008}, due to different experimental conditions and
probably also the existence of impurities.  Although still hard to measure the
friction between defect-free tubes, there is no doubt that in order to achieve
a strong bundle, tubes within it should be sufficiently long.

Here we show, from a series of molecular dynamics simulations, that
pre-pressing on CNT bundles can greatly enhance intertube friction and
consequently the strength of the nanotube fibers.  The underlying mechanism
involves the structural transition of CNTs, accompanied by an increase of
intertube frictional force and the decrease of the cross-sectional area as
well.  All tubes collapse above a critical pressure and remain collapsed after
unloading, especially the large tubes.  The friction increase, by a factor
ranging from 1.5 to 4, is chirality dependent, and is strongest for nonchiral
tubes.  Taking into account this new feature of pre-pressing, it might be
possible to spin nanotube fibers stronger than carbon fibers under current
spinning techniques.  Furthermore, although strongly related to previous
studies on radial mechanical translation during the transition of individual
CNTs or CNT bundles \cite{elliott.ja:2004, zhang.xh:20041, elliott.ja:20041,
zhang.xh:2004, sun.dy:2004, ye.x:2005, gadagkar.v:2006, ye.x:2007, zang.j:2007,
shanavas.kv:2009}, our study focusing on axial translation is obviously new and
shows inspirational results.

\section{Results and discussion}

\begin{figure}[t]
  \centering
    \includegraphics[width=2.5in]{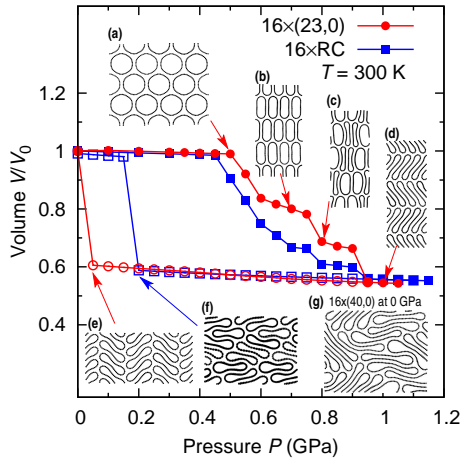}
  \caption{\label{fig.vol.vs.p}
    Volume vs pressure for the 16$\times$(23,0) and 16$\times$RC bundles. $V_0$
    is the volume at zero pressure.  The bundles show step-by-step structural
    transition, shown in filled circles and squares.  With unloading the
    pressure (open circles and squares), the bundles remain collapsed until the
    return pressure below which the bundles expand to the initial structure.
    Insects (a)--(d): Snapshots of the 16$\times$(23,0) bundle during the
    pressure loading up to 0.5, 0.7, 0.8, and 1.0 GPa, respectively.  (e):
    Collapsed structure is remained when the 16$\times$(23,0) bundle is
    unloaded down to 0.05 GPa. (f): Collapsed 16$\times$RC bundle at 0.2 GPa.
    (g): Larger tubes as the 16$\times$(40,0) remain collapsed even under zero
    pressure. Here $T = 300$ K.}
\end{figure}

The simulation is set for defect-free single-wall CNT bundles which, named by
tube number in the box and tube chirality, are 16$\times$(23,0),
16$\times$(40,0), and 16$\times$random-chirality (RC), respectively.
\ref{fig.vol.vs.p} shows the volume-pressure ($V$--$P$) relation during the
pressure loading up to 1.2 GPa and unloading down to 0 GPa for the
16$\times$(23,0) and 16$\times$RC bundles.  The bundles show step-by-step
structural transition.  We define transition pressures $P_t^{\rm start}$ and
$P_t^{\rm end}$ to denote the start and end of the transition zone.  For the
16$\times$(23,0) $P_t^{\rm start} = 0.5$ GPa and $P_t^{\rm end} = 0.95$ GPa,
and for the 16$\times$RC they are 0.45 and 0.95 GPa, respectively.  With
unloading, all tubes remain collapsed until the return pressure $P_r$, 0.05 GPa
for the 16$\times$(23,0) and 0.2 GPa the other, below which each tube expands
to the initial structure with big hollow space inside the tube.  The transition
pressures are almost the same because they are mainly radius-dependent
\cite{zhang.xh:2004}, while the return pressures are different due to the
radius inhomogeneity.  For bundles with large tube radius, for example the
16$\times$(40,0) whose $V$--$P$ curve is not shown here, $P_t^{\rm start} = 0.1$
GPa, $P_t^{\rm end} = 0.35$ GPa, and $P_r < 0$.
Therefore it is possible to get collapsed CNT bundles under atmospheric
pressure.  The clear experimental evidence of collapsed nanotubes was very
recently reported by the Windle group \cite{motta.m:2007} from the observation
of the ``dog-bone'' cross section of double- and triple-wall nanotubes with
equivalent diameter larger than $\sim$5 nm.  It presents a possible way to
improve the fibers during the direct spinning process, and it also will be of
great importance to find a similar way for the spinning out of CNT forests.
Insects (a)-(e) in \ref{fig.vol.vs.p} show the structural changes during
pressure loading and the unloaded structure for the 16$\times$(23,0) bundle.
The unloaded 16$\times$RC and 16$\times$(40,0) bundles with collapsed tubes are
shown in \ref{fig.vol.vs.p} as insects (f) and (g).  The herringbone structure
is obtained due to our large compression rate \cite{shanavas.kv:2009}.

\begin{figure}[t]
  \centering
    \includegraphics[height=2.8in,angle=-90]{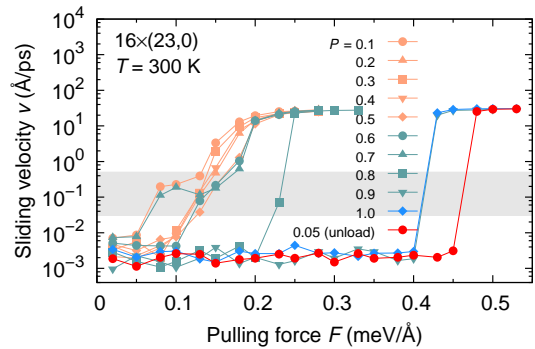}
  \caption{\label{fig.vel.vs.F}
    Sliding velocity of the sliding tube as function of the pulling force $F$
    under different loadings for the 16$\times$(23,0) bundle.  Speeds are all
    abstract values, and those smaller than 0.01 {\AA}/ps are of the same
    magnitude of the error by averaging within finite time, and thus correspond
    to zero velocity.  Below the gray zone the tube stays static, and above it
    the tube slides at a speed plateau of 2600$\sim$3000 m/s, depending on the
    pressure.}
\end{figure}

As a model study, we fix one of the tubes and pull another as far as possible
under an external pulling force $F$, along which we define the positive
direction.  We find, due to the strong commensurability of the 16$\times$(23,0)
bundle, the tube under pulling does not slide until $F$ is larger than the
so-called depinning force $f^{\rm static}$ (static friction).
\ref{fig.vel.vs.F} shows the sliding velocity at different pulling forces, each
value extracted from a simulation longer than 100 ps.  Velocities smaller than
0.01 {\AA}/ps, below the gray horizontal zone in \ref{fig.vel.vs.F}, are
considered to be at rest because such values, either positive or negative, are
within the systematic error due to the finite time average.  The tube only
slides when $F$ goes beyond the pressure-dependent depinning force which will
be discussed below, and the tube speeds up, crosses the gray zone quickly, and
finally slides at a speed plateau of 2600$\sim$3000 m/s due to phonon
excitations.  The speed plateau and its mechanism have been reported very
recently \cite{zhang.xh:2007, zhang.xh:2009}, and are beyond the scope of
current study.  Obviously, such huge sliding speed is unreal, and as a result
the bundle breaks due to the sliding.  The sliding-induced breakage has already
been observed as an abrupt diameter change in the tensile-loading experiment
\cite{yu.mf:20001}, and should be ubiquitous in fibers.  Those speeds inside
the gray zone in \ref{fig.vel.vs.F} are actually unstable as we observed from
simulation that the tube stops and slides intermittently.  More interesting and
important is $f^{\rm static}$ changes with loading.  Before the transition, it
is 0.1 meV/{\AA} per atom, and is increased to four times when collapse
happens.  With unloading, $f^{\rm static}$ varies between 0.4 and 0.45
meV/{\AA}, as the case under 0.05 GPa shown in \ref{fig.vel.vs.F}.

With considering the change of the cross-sectional area of different structures
one can estimate the increase of the tensile strength $\sigma$,
\begin{equation}
  \sigma = \sum f^{\rm static} / A,
\end{equation}
where the summation is over all atoms of the tube under pulling and $A$ is the
total cross-sectional area divided by the tube number in the simulation box.
This equation is valid because within the breaking strain (less than 10\%) of
CNT bundles \cite{yu.mf:20001} individual tubes show nearly longitudinal bond
elongation rather than bond breaking and rotation and thus still have nearly
the TPa stiffness \cite{dumitrica.t:2006}.  It means that, our assumption, the
summation of pulling forces is identical to the external load, is correct.
When the load becomes larger than $\sigma$, it can not be totally transferred
to neighboring tubes as the sliding happens.  Before the transition, {\it i.e.}
at 0.5 GPa, $A$ = 378 {\AA}$^2$ and $\sigma = 31.2$ MPa for present simulation
box length about 34 {\AA}.  To reach the strength of 10 GPa, the bundle should
be 1090 nm long, which agrees very well with the experiment \cite{yu.mf:20001}
and of the same order as a recent theoretical study \cite{qian.d:2003}.  Now
with pre-pressing to fully collapsed and with unloading down to 0.05 GPa, $A$
reduces to 232 {\AA}$^2$, $\sigma$ is improved by $4\times1.63=6.52$ due to the
quadruple increase of $f^{\rm static}$ and the ratio 1.63 between area changes.
Thus the 10 GPa tensile strength can be achieved by a bundle longer than
$1090/6.52 \approx 167$ nm because all the load can be transferred between
tubes.

\begin{figure}[t]
  \centering
    \includegraphics[height=2.8in,angle=-90]{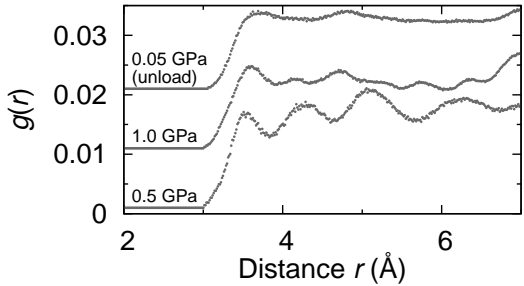}
  \caption{\label{fig.pair.dis.func}
    Partial pair distribution function $g(r)$ before nanotube collapse (0.5
    GPa), after the collapse (1.0 GPa), and after the unloading (0.05 GPa),
    of the 16$\times$(23,0) bundle.}
\end{figure}

One reason for the increase of friction is the larger tube-tube contact area
for collapsed tubes. However, it is not enough to make the quadruple friction
increase because the contact goes up much less than twice. To investigate the
structure changes, we define the partial pair distribution function (PPDF),
\begin{equation}
  g_{\alpha}(r) =
    \frac{V}{ 4\pi r^2 N_\alpha(N_{\rm tot}-N_\alpha)}
      \sum_{i=1}^{N_\alpha} \sum_{\beta\ne\alpha} \sum_{j=1}^{N_\beta}
         \delta( r - \left| \vec{r}_i - \vec{r}_j \right| ),
\end{equation}
where $\alpha$ and $\beta$ denote different tubes with atom numbers
$N_{\alpha}$ and $N_{\beta}$, respectively, $N_{\rm tot}$ the total atom
number, $V$ the volume of simulation box, $i$ and $j$ the carbon atoms of tube
$\alpha$ and $\beta$, and $\vec{r}_i-\vec{r}_j$ the displacement.  The result,
averaged among all tubes, is shown in \ref{fig.pair.dis.func}.  The first
distribution peak does not show clear changes before and after the structure
transition, {\it i.e.}, from 0.5 to 1.0 GPa, because it reflects the averaged
tube-tube distance about 3.5 {\AA} by using the current intertube potential.
There are changes in the second peak, as its position shifts from $r$ = 4.45
{\AA} to 4.25 {\AA} and the strength drops greatly after the transition.  Even
with unloading, such changes still exist except that all peaks become wider and
the total PPDF curve smoother.  The third one around $r$ = 4.75 {\AA} at 1.0
GPa is a new peak and an evidence of AB stacking between intertube
graphite-like layers.  Let us consider graphite structure and compare to the
PPDF at 1.0 GPa.  In the AB stacking with spacing $h$ , the first-, second-,
third-, and fourth-nearest distances from atoms of one layer to those of the
other, are $h$, $\sqrt{h^2+a^2}$, $\sqrt{h^2+3a^2}$, and $\sqrt{h^2+4a^2}$,
respectively, where $a$ = 1.42 {\AA} is C-C bond length.  The number ratio
between them is 1:9:6:9.  Taking $h$ = 3.5 {\AA} from our simulation, one can
find the second-, third- and fourth-nearest distances correspond exactly to the
three PPDF peaks, respectively, with proper strengths.  The nearest distance
$h$ is so close to the second one, with small number ratio as well, that the
$h$ -peak is overlapped as shown in \ref{fig.pair.dis.func}.  Here we just plot
the PPDF for $r < 7$ {\AA} in order to make clear the structural changes.  If
we extend the range to $r > 10$ {\AA}, each PPDF curve reaches a constant that
depends on the atom density.  For example, the PPDF constant at 1.0 GPa is
almost twice of that at 0.5 GPa and slightly greater than the unloaded
structure at 0.05 GPa.

\begin{figure}[t]
  \centering
    \includegraphics[width=2.8in]{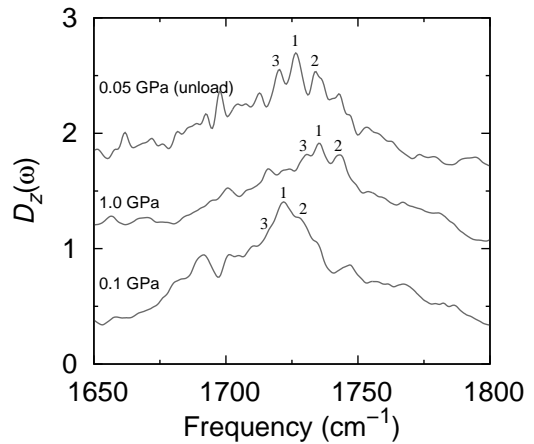}
  \caption{\label{fig.vdos}
    Vibrational density of states of the 16$\times$(23,0) bundle.  Peaks
    numbered as 1, 2 and 3, for example, shift to higher energy after the
    transition (1.0 GPa) and after unloading to 0.05 GPa.  These peaks are
    graphite-like modes, so the shift is larger at 1.0 GPa because the tube
    layers are more flattened.  The energies are higher than the experimental
    values (about 1600 cm$^{-1}$) due to the current classical atomic
    potentials.}
\end{figure}

The graphite-like structure of the collapsed CNT bundles can be also verified
from calculations of the vibrational density of states (VDOS),
\begin{equation}
  D_z(\omega) = \int e^{-i\omega t} \left< v_z(t) v_z(0) \right> dt,
\end{equation}
where $D_z(\omega)$ denotes the VDOS along the $z$ axis and $v_z(t)$ the
velocity of atoms along $z$.  \ref{fig.vdos} shows the graphite-like modes
before and after the transition and after unloading to 0.05 GPa.  Three normal
modes are labeled to show the energy shift, which correspond to the G band
shift that can be observed in experiments \cite{venkateswaran.ud:1999,
peters.mj:2000}.  The shift is larger at 1.0 GPa because the tubes are more
graphite-like, as shown in \ref{fig.vol.vs.p}.  After the unloading, it is
still detectable thus Raman scattering might be an efficient way to detect the
tube collapse.

\begin{figure}[t]
  \centering
    \includegraphics[height=2.6in,angle=-90]{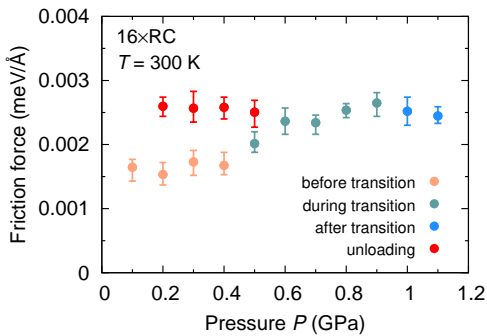}
  \caption{\label{fig.sliding.friction}
   Sliding friction under different pressures for the 16$\times$RC bundle.
   Frictions are extracted from Newton's equation, see the text.  Frictional
   force goes up by a factor of 1.5 when the transition happens and also after
   the unloading.}
\end{figure}

However, for the worst case of commensurability, the 16$\times$RC bundle, the
intertube frictional force smaller than 0.002 meV/{\AA} per carbon atom is
about two orders of magnitude smaller, in agreement with experiments
\cite{dienwiebel.m:2004, cumings.j:2000}.  Furthermore, in our pulling
simulation, it is the sliding friction rather than the depinning static force.
However, we still find the pre-pressing also increases the sliding friction
between tubes.  The tube under pulling starts to move from the optimized
structure when $F$ = 0.005 meV/{\AA} is applied, either in or out along the
tube axis $z$.  The friction $f$ is extracted from Newton's equation when the
sliding speed of that tube increases from zero to around 1 {\AA}/ps.  For
example, at 0.1 GPa, the speed goes up linearly from zero to 0.8 {\AA}/ps in
300 ps, corresponding to an acceleration of $a = 0.00267$ {\AA}/ps$^2$.  The
sliding friction is $f = ma - F =-0.00168$ meV/{\AA}, opposite to the pulling
direction.  However, after unloading to 0.2 GPa, it takes more than 420 ps to
reach the speed of 0.8 {\AA}/ps from zero, and the friction is extracted as
-0.00263 meV/{\AA}.  In \ref{fig.sliding.friction}, we plot the friction forces
under different pressures where error bars indicate the friction fluctuation
among several pulling simulations.  Before the transition, frictional forces
are all around 0.0017 meV/{\AA}.  It goes up greatly to 0.0026 meV/{\AA}, by a
factor of 1.5, when the transition happens.  The friction value maintains after
the transition and also after the unloading.  Taking into account the
cross-sectional area changes, the strength of a bundle composed of tubes with
random chirality is almost tripled.

\section{Conclusion}

In summary, we have investigated the mechanical axial translation between
nanotubes within CNT bundles.  With a treatment of pre-pressing, the collapsed
and compact structure remained under a pressure as small as comparable to
atmospheric pressure.  After the collapse of tubes, not only the
cross-sectional area decreases, but the tube-tube frictional force also goes
up, especially by a factor of about 4 if the tubes are commensurate.  Our study
suggests a new and efficient way to reinforce the strength of CNT fibers, and
actually we have already been on the avenue of experimental studies.

\section{Methods}

\begin{figure}[t]
  \centering
    \includegraphics[width=2.3in]{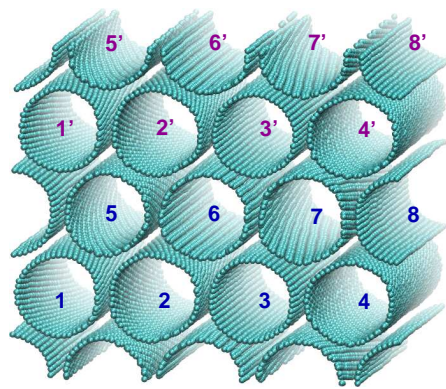}
  \caption{\label{fig.tube-indexing}
    Tube arrangement of the 16$\times$RC bundle in the simulation box.  Primed
    tubes are copies of those unprimed.  Tube 1 is fixed to move while tube
    $3'$ is driven under an external pulling force acting on each atom.  All
    other tubes are free to move.}
\end{figure}

\begin{table}
  \caption{\label{tab.tube-details} Tube chirality, radius ($R$), length ($L$),
    and number of atoms ($N$) in present simulation.  Note that $L$ changes
    slightly in simulation due to the barostat.}
  \begin{tabular}{ccccc}
    \hline
    Tube                  & Chirality & $R$ ({\AA}) & $L$ ({\AA})  & $N$ \\
    \hline
    16$\times$(40,0): 1-8 & (40,0)    & 15.66       &  21.30       &  800 \\
    16$\times$(23,0): 1-8 & (23,0)    &  9.00       &  34.08       &  736 \\
    16$\times$RC: 1       & (14,14)   &  9.49       & 104.00       & 2240 \\
    16$\times$RC: 2       & (16,12)   &  9.52       & 104.00       & 2368 \\
    16$\times$RC: 3       & (18,8)    &  9.03       & 104.00       & 2128 \\
    16$\times$RC: 4       & (20,6)    &  9.23       & 104.00       & 2224 \\
    16$\times$RC: 5       & (22,2)    &  9.03       & 104.00       & 2128 \\
    16$\times$RC: 6       & (22,4)    &  9.49       & 104.00       & 2352 \\
    16$\times$RC: 7       & (23,0)    &  9.00       & 104.00       & 2208 \\
    16$\times$RC: 8       & (24,0)    &  9.39       & 104.00       & 2304 \\
    \hline
  \end{tabular}
\end{table}

Periodic boundary conditions are used in all three dimensions.  The tubes are
initially assembled in hexagonal symmetry, with indexing numbers for the
16$\times$RC bundle shown in \ref{fig.tube-indexing} where primed tubes
have the same chiralities as those unprimed.
We list the chirality, radius, length, and atom number of each tube in
\ref{tab.tube-details}.  The length is sufficiently long to
effectively reflect the energy transfer between tubes and to
avoid the size effect where self-diffusion can be caused for short tubes due to
the small energy variation between neighboring tubes.  One constraint is used
to fix the center of mass of tube 1 (see \ref{fig.tube-indexing}), while tube
$3'$ is driven to move along tube axis $z$ under an external pulling force $F$
acting on each atom.  We extract the static frictional force $f^{\rm static}$,
if possible, by assigning the critical pulling force below which no tube
sliding happens and slides otherwise.  The sliding frictional force $f$ is
extracted from the displacement-time curve of tube $3'$ by using Newton's
equation $F + f = ma$, $m$ being the mass of carbon atom and $a$ the overall
acceleration along $z$.  Temperature $T = 300$ K and pressure are globally
controlled by the Berendsen's algorithm \cite{berendsen.hjc:1984}. We set along
$z$ zero pressure, $P_z = 0$, while the cross-sectional pressure $P_x = P_y$
varies between 0 and 1.2 GPa.  The homogeneous isothermal compressibility in
the cross-sectional plane is chosen to be one order of magnitude larger than
that along $z$.  The intratube C-C covalent bonds are described by the reactive
empirical bond-order potential \cite{brenner.dw:2002}, while the intertube van
der Waals interactions and those between intratube non-neighbor atoms by a
Lennard-Jones potential \cite{mao.zg:1999}, as our previous study has used
\cite{zhang.xh:2004}.

\begin{acknowledgements}
We acknowledge funding support from Chinese Academy of Science (hundred talent
program, and knowledge innovation program), and international collaboration
project by Ministry of Science and technology. We also thank Prof. Yuntian Zhu
for instructive comments. Parts of the calculations were performed in the
Shanghai Supercomputer Center (SSC) of China.
\end{acknowledgements}

%



\end{document}